\documentclass[10pt,aps,prd,reprint,superscriptaddress,longbibliography,nofootinbib,floatfix]{revtex4-2}
\usepackage{mathrsfs,amsmath,amsthm,latexsym,amssymb,amsfonts,graphicx,txfonts}
\usepackage[utf8]{inputenc}
\usepackage[T1]{fontenc}
\usepackage[colorlinks=true,linkcolor=blue,citecolor=blue,urlcolor=blue]{hyperref}
\usepackage{orcidlink}
\usepackage{booktabs}
\allowdisplaybreaks
\newcommand{\ABTU}{Department of Physics, Al-Hussein Bin Talal University, 71111, Ma'an, Jordan}
\newcommand{\RGUI}{Department of Physics, The Assam Royal Global University, Guwahati-781035, Assam, India}
\newcommand{\UFPB}{Departamento de Física, Universidade Federal da Paraíba, Centro de Ciências Exatas e da Natureza, 58051-970, João Pessoa, Paraíba, Brazil}
\newcommand{\UCCB}{Programa de P\'os-Gradua\c c\~ao em F\'{\i}sica \& Coordena\c c\~ao do Curso de F\'{\i}sica -- Bacharelado, Universidade Federal do Maranh\~{a}o, 65085-580 S\~{a}o Lu\'{\i}s, Maranh\~{a}o, Brazil}

\begin{document}
\baselineskip=12pt

\title{Transfer observables of rotating acoustic black holes from ray tracing: shadow centroid, redshift asymmetry and flux imbalance}

\author{Faizuddin Ahmed\orcidlink{0000-0003-2196-9622}}
\email{faizuddinahmed15@gmail.com}
\affiliation{\RGUI}
\author{Ahmad Al-Badawi\orcidlink{0000-0002-3127-3453}}
\email{ahmadbadawi@ahu.edu.jo }
\affiliation{\ABTU}
\author{Fernando M. Belchior\orcidlink{0009-0006-8675-7849}}
\email{fernandobelcks7@gmail.com}
\affiliation{\UFPB}
\author{Edilberto O. Silva\orcidlink{0000-0002-0297-5747}}
\email{edilberto.silva@ufma.br}
\affiliation{\UCCB}

\begin{abstract}
We construct an impact-parameter-resolved transfer framework for null acoustic rays in the rotating draining-bathtub spacetime.  The formalism separates the source-independent ray geometry from the source and detector model by keeping explicit the acoustic redshift, transfer convention, emissivity, emitter velocity field, and source-to-screen mapping.  The geometric capture interval provides two clean observables: a shadow centroid that shifts linearly with circulation and a shadow width that grows monotonically with circulation.  Observable profiles are obtained from direct ray-source intersections, finite source width or extended-disk integration, detector convolution, and convergence checks, rather than from an approximate semi-analytic ring map.  The transfer calculation shows that rotation produces a left-right redshift tilt and a branch-dependent flux imbalance, while the total flux alone remains a degenerate circulation diagnostic.  The most useful diagnostics are differential quantities: the shadow centroid, branch-integrated flux asymmetry, peak asymmetry, left-right redshift asymmetry, and global redshift contrast.  We also discuss how these observables respond to the transfer convention, intrinsic azimuthal emissivity, the choice of left-right split, finite resolution, and physical limitations such as dispersion, viscosity, and finite-depth corrections.
\end{abstract}

\maketitle

\section{Introduction}\label{sec:introduction}

Black-hole images are not determined by the capture region alone.  The dark
depression associated with null capture provides the geometric skeleton of the
shadow, but the observed brightness pattern also depends on redshift, Doppler
weighting, lensing, the velocity field of the emitting material, and the
intrinsic emissivity of the source.  This distinction is already present in the
classic ray-tracing calculations of compact-object appearance
\cite{Luminet1979,CunninghamBardeen1972}, in modern analyses of photon rings
and lensing rings \cite{CunhaHerdeiro2018,Gralla2019,Gralla2020,Johnson2020},
and in the observational interpretation of the Event Horizon Telescope images
of M87$^*$ and Sgr~A$^*$ \cite{EHT2019I,EHT2019II,EHT2019III,EHT2019IV,EHT2019V,EHT2019VI,EHT2022I,EHT2022VI}.
For the present purpose, the important lesson is methodological rather than
astrophysical: a shadow observable should be formulated as a transfer problem,
not only as a critical-impact-parameter calculation.

Analogue gravity offers a controlled setting in which this separation between
geometry and transfer can be tested.  Sound waves in transonic flows experience
effective horizons and curved acoustic geometries \cite{Unruh1981,Visser1998},
and analogue systems have been used to study Hawking-like emission,
mode conversion, black-hole lasing, entanglement, and hydrodynamic
superradiance \cite{Lahav2010,Philbin2008,Rousseaux2008,Weinfurtner2011,Steinhauer2014,Steinhauer2016,MunozdeNova2019,Kolobov2021,Torres2017}.
The rotating draining-bathtub flow is especially relevant because it contains
an acoustic horizon generated by radial draining and an ergoregion generated
by circulation.  Its perturbations exhibit quasinormal ringing, late-time
tails, absorption, orbiting-type scattering, resonances, and superradiant
amplification \cite{Berti2004,Dolan2009,Oliveira2010,Dolan2012,Dolan2013,Anacleto2011,Cardoso2004b,Patrick2018,Berti2009,Brito2015}.
These results show that the draining-bathtub geometry is a useful analogue of
rotating black-hole physics, but most previous work has focused on waves,
scattering, spectra, or amplification rather than on the transferred acoustic
image itself.

The question addressed here is therefore more specific: how does the
circulation of a rotating acoustic vortex appear in an impact-parameter
resolved acoustic shadow, redshift profile, and flux profile?  We formulate
the answer as a ray-tracing transfer calculation.  Null acoustic rays are
traced from the observer screen to the emitting region; real intersections
with rings and extended disks are computed; the acoustic redshift is evaluated
on each branch; and the source-to-screen mapping is included explicitly.
The emitter may rotate with an arbitrary angular velocity
$\Omega_{\rm em}(r)$, may have radial velocity $v^r_{\rm em}(r)$, and may have
a nonaxisymmetric emissivity $\mathcal I_{\rm em}(r,\phi)$.  The transfer
exponent $\eta$ in $I_{\rm obs}\propto g_{\rm ac}^{\eta}$ is kept explicit
because the appropriate power depends on the acoustic observable and detector
response.

The formulation is intended as an effective-metric analogue of shadow-based
rotation diagnostics rather than as a hydrodynamic model of one particular
apparatus.  In this sense the calculation plays a role similar to relativistic
ray tracing in gravitational black-hole imaging: the null capture region fixes
a source-independent skeleton, while transfer, emitter motion, and detector
response determine which part of that skeleton is visible in an intensity
profile.  The draining-bathtub system is useful precisely because the
circulation is tunable and the separation between geometric observables and
transfer-dependent observables can be displayed analytically and numerically.

The main physical point is that rotation is most cleanly identified through
differential observables.  The circulation parameter $B$ shifts the critical
impact parameters and hence the acoustic shadow interval.  It also produces a
left-right redshift imbalance and a corresponding brightness asymmetry.
However, a total integrated flux can be insensitive or degenerate because
rotation may redistribute intensity between the two sides of the image.  We
therefore focus on branch-integrated flux asymmetry, peak asymmetry,
left-right redshift asymmetry, and global redshift contrast.  These
quantities are not claimed to be source-independent in all circumstances;
rather, they provide a controlled hierarchy of observables once the source
symmetry, detector resolution, and transfer convention are specified.

\begin{widetext}
\begin{center}
\refstepcounter{table}\label{tab:contributions}
\begin{minipage}{0.96\textwidth}
\textbf{TABLE~\thetable.} Main additions of the present transfer framework relative to a purely geometric or semi-analytic ring prescription.
\end{minipage}
\vspace{0.5em}
\begin{ruledtabular}
\begin{tabular}{lll}
Ingredient & Role in the calculation & Why it matters observationally \\
\hline
Analytic shadow interval & Gives $b_c^\pm$, centroid and width & Separates geometry from source transfer \\
Real ray-source intersections & Replaces approximate emission angles & Controls branch structure and caustics \\
Explicit acoustic redshift & Evaluates $g_{\rm ac}$ on each ray branch & Connects circulation with frequency shift \\
Transfer exponent $\eta$ & Encodes the measured acoustic intensity convention & Allows detector-dependent calibration \\
Jacobian/source width & Regularizes thin-ring caustics & Prevents singular ring artifacts from dominating \\
Direct disk integration & Produces finite-resolution profiles & Gives the main observables used in the paper \\
Differential diagnostics & Uses branch fluxes and redshift asymmetries & Reduces dependence on total normalization
\end{tabular}
\end{ruledtabular}
\end{center}
\end{widetext}

The paper is organized as follows.  Section~\ref{sec:geometry} introduces the
draining-bathtub geometry and derives the acoustic shadow interval.  Section
\ref{sec:transfer} develops the redshift and intensity-transfer prescription,
including normalization and numerical conventions.  Section~\ref{sec:results}
presents the ray-traced profiles, differential diagnostics, regularization
study, convergence test, and robustness considerations.  Section
\ref{sec:conclusions} summarizes the implications for transfer diagnostics in
rotating acoustic effective metrics.

\section{Geometry, null rays and acoustic shadow}
\label{sec:geometry}

We consider the $(2+1)$-dimensional draining-bathtub acoustic black hole.
In laboratory-type polar coordinates, the effective metric can be written as
\begin{equation}
 ds^2=dt^2-\left(dr-\frac{A}{r}dt\right)^2-\left(r d\phi-\frac{B}{r}dt\right)^2,
 \label{eq:lab_metric}
\end{equation}
where $A$ controls the radial draining flow and $B$ controls the circulation.
The radial flow determines the sonic horizon; the circulation determines the
rotational dragging of acoustic rays. Introducing Boyer--Lindquist-like
coordinates,
\begin{equation}
 dt=d\tilde t-\frac{A r}{r^2-A^2}dr,
 \qquad
 d\phi=d\tilde\phi-\frac{AB}{r(r^2-A^2)}dr,
 \label{eq:bl_transform}
\end{equation}
we obtain
\begin{equation}
 ds^2=g(r)d\tilde t^{2}-\frac{dr^2}{f(r)}+2B\,d\tilde t\,d\tilde\phi-r^2 d\tilde\phi^{2},
 \label{eq:bl_metric}
\end{equation}
with
\begin{equation}
 f(r)=1-\frac{A^2}{r^2},
 \qquad
 g(r)=1-\frac{A^2+B^2}{r^2}.
 \label{eq:fg_functions}
\end{equation}
The acoustic horizon is located at
\begin{equation}
 r_h=A,
 \label{eq:horizon}
\end{equation}
where $f(r_h)=0$, while the ergosurface is located at
\begin{equation}
 r_e=\sqrt{A^2+B^2},
 \label{eq:ergosurface}
\end{equation}
where $g(r_e)=0$. In what follows we drop the tildes. The nonzero components
are
\begin{equation}
 g_{tt}=g(r),\;\;\; g_{t\phi}=B,\;\;\;
 g_{rr}=-f(r)^{-1},\;\;\; g_{\phi\phi}=-r^2.
 \label{eq:metric_components}
\end{equation}

The acoustic shadow is determined by null rays. The null condition is
\begin{equation}
 g_{\mu\nu}\dot x^\mu\dot x^\nu=0,
 \label{eq:null_condition}
\end{equation}
where a dot denotes differentiation with respect to an affine parameter.
Stationarity and axial symmetry give the conserved quantities
\begin{equation}
 E=g(r)\dot t+B\dot\phi,
 \qquad
 L=r^2\dot\phi-B\dot t.
 \label{eq:conserved_quantities}
\end{equation}
With this convention the covariant angular momentum is $k_\phi=-L$, so that
the impact parameter used below is $b=L/E$.
Solving for the velocities gives
\begin{equation}
 \dot t=\frac{E-BL/r^2}{f(r)},
 \label{eq:tdot}
\end{equation}
\begin{equation}
 \dot\phi=\frac{L}{r^2}+\frac{BE-LB^2/r^2}{r^2 f(r)}.
 \label{eq:phidot}
\end{equation}
Substitution in the null condition yields
\begin{equation}
 \frac{\dot r^2}{E^2}=1-\frac{b^2 g(r)+2Bb}{r^2}
 \equiv 1-V_{\rm eff}(r;b),
 \label{eq:radial_b}
\end{equation}
where $b=L/E$ is the impact parameter and
\begin{equation}
 V_{\rm eff}(r;b)=\frac{b^2 g(r)+2Bb}{r^2}.
 \label{eq:effective_potential}
\end{equation}
Critical rays satisfy $\dot r=0$ and $d\dot r^2/dr=0$. In the dimensionless
parametrization $\beta=B/A$, the corresponding impact parameters are
\begin{equation}
    \frac{b_c^{\pm}}{A}=-2\beta\pm2\sqrt{1+\beta^2}.
    \label{eq:critical_impact_explicit}
\end{equation}
The shadow interval is the set of observer-screen directions associated with
captured rays. For a distant observer in the equatorial $(2+1)$ geometry, the
screen coordinate may be identified with the asymptotic impact parameter,
$X/A\simeq b/A$.

It is useful to define the dimensionless midpoint and width
\begin{equation}
 \bar b_{\rm mid}\equiv\frac{b_c^++b_c^-}{2A},
 \qquad
 \Delta\bar b\equiv\frac{b_c^+-b_c^-}{A}.
 \label{eq:shadow_dimensionless_definitions}
\end{equation}
They follow directly from Eq.~\eqref{eq:critical_impact_explicit}:
\begin{equation}
 \bar b_{\rm mid}=-2\beta,
 \qquad
 \Delta\bar b=4\sqrt{1+\beta^2}.
\label{eq:shadow_midpoint_width}
\end{equation}
Equivalently, in dimensionful form,
\begin{equation}
 b_{\rm mid}\equiv\frac{b_c^++b_c^-}{2}=-2B,
 \qquad
 \Delta b=b_c^+-b_c^-=4A\sqrt{1+\beta^2}.
 \label{eq:shadow_midpoint_width_dimensionful}
\end{equation}
The centroid shift is exact and \emph{linear} in $\beta$; the width grows
monotonically as $\sqrt{1+\beta^2}$. In the two physically important limits,
\begin{align}
  \beta\ll 1:\quad &b_{\rm mid}\approx -2\beta A=-2B,\quad
  \Delta b\approx 4A\!\left(1+\tfrac{1}{2}\beta^2\right),
  \label{eq:shadow_smallbeta}\\
  \beta\gg 1:\quad &b_{\rm mid}\approx -2\beta A=-2B,\quad
  \Delta b\approx 4\beta A=4B.
  \label{eq:shadow_largebeta}
\end{align}
At small circulation, the centroid shifts at first order in $B/A$ while the
width increases only at second order; at large circulation, both the centroid
and the width grow linearly with $B/A$, so measuring both quantities in a
single experiment provides two independent constraints on $A$ and $B$.
These analytic results are the acoustic counterpart of the spin-dependent
photon-capture range in Kerr gravity and do not depend on any source model.

\section{General redshift and intensity-transfer prescription}
\label{sec:transfer}

The acoustic frequency measured by an observer with velocity $u^\mu$ is
\begin{equation}
 \omega_{\rm meas}=k_\mu u^\mu,
 \label{eq:measured_frequency}
\end{equation}
where $k^\mu=\dot x^\mu$ is the tangent to the null acoustic ray and $k_\mu=g_{\mu\nu}k^\nu$. With the mostly-minus signature used here, and with future-directed acoustic rays normalized by positive $E=k_t$, we use $\omega_{\rm meas}=k_\mu u^\mu$. The redshift
factor is therefore
\begin{equation}
 g_{\rm ac}=\frac{\omega_{\rm obs}}{\omega_{\rm em}}
 =\frac{(k_\mu u^\mu)_{\rm obs}}{(k_\mu u^\mu)_{\rm em}}.
 \label{eq:redshift_general}
\end{equation}
For a static observer at large radius,
$(k_\mu u^\mu)_{\rm obs}\simeq E$. The emitter is taken to have a general
velocity field
\begin{equation}
 u^\mu_{\rm em}=u^t_{\rm em}\left(1,v^r_{\rm em},\Omega_{\rm em}\right),
 \label{eq:uem_general}
\end{equation}
where $\Omega_{\rm em}=d\phi/dt$ is the angular velocity and
$v^r_{\rm em}=dr/dt$ is an optional radial coordinate velocity. The
normalization condition gives
\begin{equation}
 \left(u^t_{\rm em}\right)^{-2}=
 g(r)+2B\Omega_{\rm em}-r^2\Omega_{\rm em}^2
 -\frac{(v^r_{\rm em})^2}{f(r)}.
 \label{eq:ut_general}
\end{equation}
Using $k_t=E$, $k_\phi=-L=-bE$, and keeping the radial contribution explicit,
the general redshift factor can be written as
\begin{equation}
 g_{\rm ac}=\frac{1}{u^t_{\rm em}\left[1-b\Omega_{\rm em}+(k_r/E)v^r_{\rm em}\right]}.
 \label{eq:gac_general}
\end{equation}
This equation is one of the central ingredients of the paper. The sign of $k_r/E$ is branch-dependent and is fixed by the traced ray; its explicit form is given in Appendix~\ref{app:inverse_metric}. It separates the
angular Doppler term $b\Omega_{\rm em}$ from the radial term
$(k_r/E)v^r_{\rm em}$, while the normalization $u^t_{\rm em}$ contains the
acoustic gravitational redshift and the frame-dragging contribution
$2B\Omega_{\rm em}$. For circular emitters, $v^r_{\rm em}=0$, and
Eq.~\eqref{eq:gac_general} reduces to
\begin{equation}
 g_{\rm ac}=\frac{\sqrt{g(r)+2B\Omega_{\rm em}-r^2\Omega_{\rm em}^2}}{1-b\Omega_{\rm em}}.
 \label{eq:gac_circular}
\end{equation}
A vortex-comoving source, $\Omega_{\rm em}=B/r^2$, is then only one possible
choice. Static emitters, constant angular velocity emitters, power-law profiles
$\Omega_{\rm em}=\Omega_0 r^{-q}$, inflow and outflow all fit into the same
formula.

The observed intensity is expressed as a transfer relation rather than as a
fixed model. If a ray labelled by screen coordinates $(X,Y)$ intersects the
source one or more times, the observed acoustic intensity is written as
\begin{equation}
 I_{\rm obs}(X,Y)=\sum_n g_{{\rm ac},n}^{\eta}\,
 \mathcal I_{\rm em}(r_n,\phi_n)\,\mathcal J_n^{-1}.
 \label{eq:iobs_general_sum}
\end{equation}
The index $n$ labels different intersections with the source, $\eta$ is a
transfer exponent, and $\mathcal J_n$ denotes the local mapping factor between
the emitting region and the observer screen.

\subsection{Transfer exponent and measured acoustic intensity}
\label{subsec:eta}

The exponent $\eta$ is kept as a phenomenological but controlled part of the
transfer law.  In photon radiative transfer in $(3+1)$ dimensions, invariance
of $I_\nu/\nu^3$ motivates $\eta=3$ for specific intensity.  For an ideal
two-dimensional acoustic field, the corresponding energy-flux scaling is closer
to $\omega^2$, which suggests $\eta=2$ as the natural acoustic reference value.
A laboratory measurement, however, need not record this idealized quantity
directly.  A microphone, surface-elevation probe, phase-imaging reconstruction,
or Fourier-filtered intensity map may respond to pressure amplitude, wave
energy density, frequency-resolved power, or a processed combination of these
quantities.  The effective power of $g_{\rm ac}$ can therefore depend on the
detector response and on the reconstruction protocol.

For this reason, the formalism does not require a universal value of $\eta$.
The fiducial numerical examples use $\eta=3$ to facilitate comparison with the
standard relativistic transfer prescription and to provide a conservative
high-contrast reference.  If $\eta=2$ is adopted, the same redshift field
produces weaker intensity contrast, but the hierarchy of diagnostics is
unchanged: the left-right redshift asymmetry is fixed by the ray kinematics,
whereas brightness asymmetries scale with the detector-dependent transfer
weight.  In an experiment, $\eta$ should be calibrated by propagating a known
source through a controlled background flow, or by matching the reconstructed
intensity to the acoustic energy convention used by the detector.

The emissivity is written in the general separable form
\begin{equation}
 \mathcal I_{\rm em}(r,\phi)=I_0\,\mathcal R(r)\,\mathcal P(\phi).
 \label{eq:emissivity_general}
\end{equation}
For an extended disk we use
\begin{equation}
 \mathcal R(r)=\left(\frac{r}{r_h}\right)^{-p}
 \Theta(r-r_{\rm in})\Theta(r_{\rm out}-r),
 \label{eq:radial_disk}
\end{equation}
while a thin ring is recovered in the limit
$\mathcal R(r)\rightarrow \delta(r-r_s)$. Nonaxisymmetric emission is
represented by
\begin{equation}
 \mathcal P(\phi)=1+\sum_{m=1}^{m_{\rm max}}\epsilon_m
 \cos[m(\phi-\phi_m)].
 \label{eq:angular_emissivity}
\end{equation}
This angular modulation describes hot spots or azimuthal density variations.
The point of Eqs.~\eqref{eq:iobs_general_sum}--\eqref{eq:angular_emissivity}
is that geometric rotation, source rotation, radial motion and intrinsic
emissivity asymmetry can be varied independently. This is essential for
identifying which features of the observed image are robust signatures of
hydrodynamic frame dragging.

A nonaxisymmetric $\mathcal P(\phi)$ can mimic part of the observed brightness
imbalance.  This is why the redshift diagnostics and the geometric shadow
centroid are kept separate from the flux diagnostics.  In the axisymmetric
fiducial model, a nonzero $\mathcal A_g^{\rm LR}$ is a direct consequence of
circulation.  With intrinsic hot spots or azimuthal density gradients, however,
$\mathcal A_I^{\rm flux}$ and $\mathcal A_I^{\rm peak}$ must be interpreted
jointly with the redshift map and with the known source modulation.  The
diagnostic strategy is therefore not to infer $B$ from a single brightness
ratio, but to seek consistency among the shadow centroid, the sign and size of
the redshift tilt, and the branch-resolved flux imbalance.

\subsection{Numerical implementation, direct ray tracing, and normalization conventions}
\label{subsec:numerical_implementation}

The numerical calculations are performed in dimensionless units with $A=1$.
The metric functions are therefore
\begin{equation}
 f(r)=1-\frac{1}{r^2},
 \qquad
 g(r)=1-\frac{1+B^2}{r^2},
 \label{eq:numerical_fg}
\end{equation}
with horizon radius $r_h=1$ and ergosurface radius $r_e=\sqrt{1+B^2}$.
The shadow boundary is evaluated analytically from
Eq.~\eqref{eq:critical_impact_explicit}; it is not inferred from finite
resolution ray counting.  This ensures that the geometric shadow interval is
kept separate from source-dependent transfer effects.

For the ray tracing, a ray is labelled by the asymptotic screen coordinate
$X/A\simeq b/A$.  The radial equation is written as
\begin{equation}
 F(r;b,B)\equiv\frac{\dot r^2}{E^2}
 =1-\frac{b^2g(r)+2Bb}{r^2},
 \label{eq:numerical_F}
\end{equation}
and the angular trajectory is obtained from
\begin{equation}
 \frac{d\phi}{dr}=\frac{\dot\phi/E}{\pm\sqrt{F(r;b,B)}} ,
 \label{eq:numerical_dphidr}
\end{equation}
where the minus sign denotes the inward branch and the plus sign denotes the
outward branch.  Turning points are found by solving
\begin{equation}
 u^2-(b^2+2Bb)u+b^2(1+B^2)=0,
 \qquad u=r^2,
 \label{eq:turning_quadratic}
\end{equation}
and retaining roots outside the horizon.  If no such root exists, the ray is
captured.  If a turning point exists, the trajectory is continued on the
outward branch.  Intersections with the source are then detected directly from
the traced trajectory, rather than by imposing the approximate relation
$\phi_{\rm em}\simeq\arcsin(b/r_s)$.

For a thin ring at $r=r_s$, each real intersection gives an emission angle
$\phi_{\rm em}(b)$ and a redshift factor $g_{\rm ac}$ computed from
Eq.~\eqref{eq:gac_general}.  The local mapping between the ring and the screen
is encoded in the one-dimensional Jacobian
\begin{equation}
 \mathcal J_{\rm ring}^{-1}(b)
 =r_s\left|\frac{d\phi_{\rm em}}{db}\right|.
 \label{eq:ring_jacobian}
\end{equation}
The derivative is computed numerically on the screen grid for each branch.
Near caustics the thin-ring Jacobian can become very large, as expected for an
infinitesimal source.  We therefore use the raw thin-ring result only as a
diagnostic limiting case.  The profiles used for physical interpretation are
regularized either by applying a smooth cap to $\mathcal J^{-1}$ or, more
physically, by replacing the delta-function ring with a finite-width Gaussian
ring,
\begin{equation}
 \mathcal R_{\rm ring}(r)=
 \exp\!\left[-\frac{(r-r_s)^2}{2\sigma_r^2}\right],
 \label{eq:finite_ring_profile}
\end{equation}
followed by a detector convolution in $X$ that preserves the integrated flux.

For the extended-disk results, we do not stack only delta-function ring
profiles.  Instead, the observed intensity is obtained by direct optically thin
integration along the traced rays,
\begin{equation}
 I_{\rm obs}(X)=
 \sum_{\rm branches}
 \int_{\lambda_{\rm in}}^{\lambda_{\rm out}}
 g_{\rm ac}^{\eta}
 \mathcal I_{\rm em}\!\left[r(\lambda),\phi(\lambda)\right]
 W[r(\lambda)]\,d\lambda ,
 \label{eq:direct_disk_integral}
\end{equation}
with $r_{\rm in}\le r(\lambda)\le r_{\rm out}$ and zero contribution outside
the disk.  In the fiducial disk runs we take $r_{\rm in}/A=2$,
$r_{\rm out}/A=8$, $\mathcal R(r)\propto r^{-p}$ with $p=2$, and
$\eta=3$.  The detector convolution width is $\sigma_X/A=0.12$, the finite-ring
width used for comparison is $\sigma_r/A=0.15$, and the fiducial screen and
radial resolutions are $N_b=260$ and $N_r=1300$.  These numbers are not fit
parameters; they define the resolution and smoothing convention used to turn
the geometric-acoustics ray map into finite-resolution observables.

The differential observables used below are defined as follows.  The
branch-resolved fluxes are
\begin{equation}
 F_\pm=\int_{\pm X>0} I_{\rm obs}(X)\,dX ,
 \label{eq:branch_fluxes}
\end{equation}
where the fiducial split is taken with respect to the laboratory screen origin.
Because the shadow centroid itself is displaced when $B\ne0$, an alternative
source-centred or shadow-centred split may be defined by replacing $X$ with
$X-b_{\rm mid}$.  The two conventions answer slightly different questions:
the laboratory split measures the directly observed left-right imbalance,
whereas the centroid-corrected split measures the residual imbalance after the
pure geometric translation has been removed.  Unless stated otherwise, the
figures use the laboratory split, while the centroid-corrected definition is
the natural choice for experiments in which the shadow displacement is first
fitted and subtracted.

The flux asymmetry is
\begin{equation}
 \mathcal A_I^{\rm flux}
 =
 \frac{|F_+-F_-|}{F_++F_-}.
 \label{eq:AI_flux}
\end{equation}
The peak asymmetry,
\begin{equation}
 \mathcal A_I^{\rm peak}
 =
 \frac{|I_{\rm max}^{(+)}-I_{\rm max}^{(-)}|}
 {I_{\rm max}^{(+)}+I_{\rm max}^{(-)}} ,
 \label{eq:AI_peak}
\end{equation}
is also shown, but it is more sensitive to caustics and resolution than
$\mathcal A_I^{\rm flux}$.  The left-right redshift asymmetry is defined from
intensity-weighted averages,
\begin{equation}
 \bar g_\pm=
 \frac{\int_{\pm X>0} g_{\rm ac}(X)I_{\rm obs}(X)dX}
 {\int_{\pm X>0} I_{\rm obs}(X)dX},
 \label{eq:gbar_pm}
\end{equation}
as
\begin{equation}
 \mathcal A_g^{\rm LR}
 =
 \frac{|\bar g_+-\bar g_-|}{\bar g_++\bar g_-}.
 \label{eq:Ag_LR}
\end{equation}
This quantity vanishes at $B=0$ by symmetry.  We also display the global
redshift contrast $\mathcal C_g$, computed from intensity-weighted quantiles of
the redshift distribution.  Unlike $\mathcal A_g^{\rm LR}$, $\mathcal C_g$ need
not vanish at $B=0$, because a nonrotating extended disk can still contain
emission from radii with different gravitational/acoustic redshifts.

\section{Results and physical interpretation}
\label{sec:results}

We now discuss the numerical results, emphasizing which features are geometric
and which require source transfer.  The original semi-analytic ring profiles
and strip maps have been replaced by direct ray-traced disk observables.  The
only figures retained from the earlier version are the two purely geometric
diagnostics: the null-ray structure and the analytic shadow interval.  These
remain useful because they are independent of the source model and of the
regularization used for the intensity.

\begin{figure}[tbhp]
    \centering
    \includegraphics[width=0.43\textwidth]{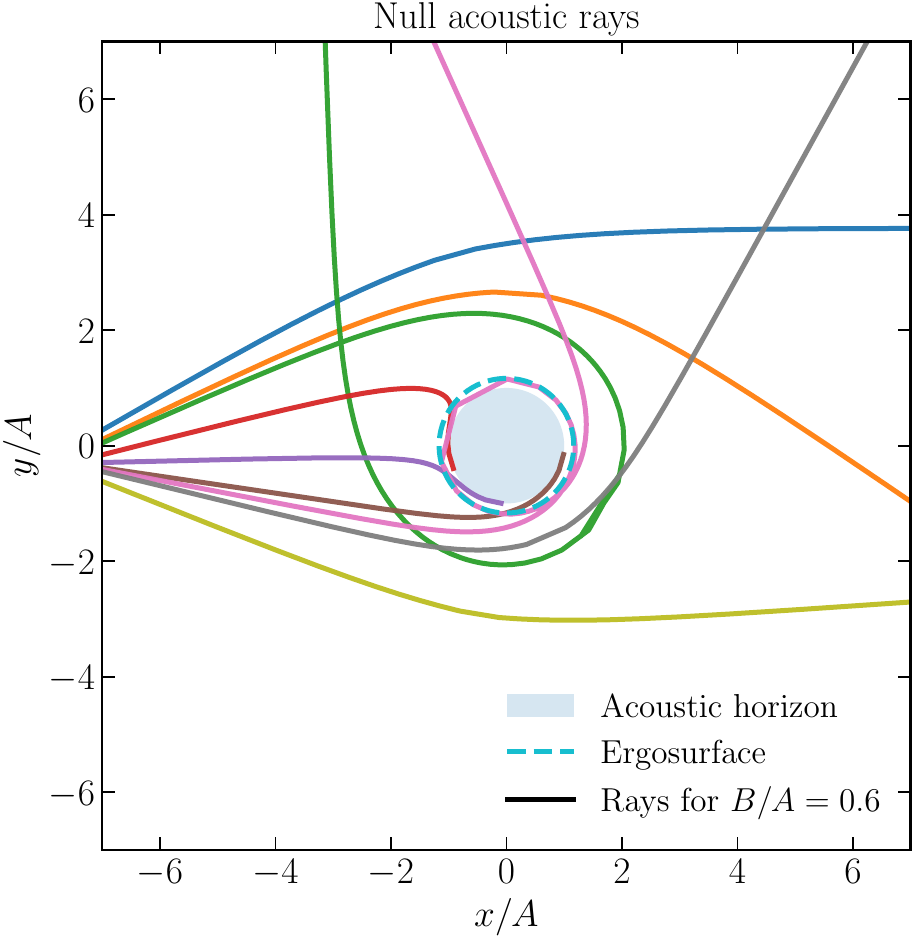}
    \caption{Null acoustic rays in the rotating draining-bathtub geometry for
    $B/A=0.6$. The filled central disk denotes the acoustic horizon $r_h=A$,
    while the dashed circle represents the ergosurface
    $r_e=\sqrt{A^2+B^2}$. Rays inside the critical interval are captured by the
    horizon, whereas rays outside it are scattered. Near-critical trajectories
    wind around the vortex before either escaping or being absorbed.}
    \label{fig:null_rays}
\end{figure}

Figure~\ref{fig:null_rays} shows the optical skeleton of the problem.  Rays
with impact parameters inside the critical interval are absorbed by the
acoustic horizon, while those outside it are scattered.  Near the critical
values the trajectories spend a long affine time around the vortex.  This is
the acoustic analogue of the photon-ring mechanism in gravitational black-hole
imaging.  In the draining-bathtub case, however, the circulation parameter is a
directly tunable property of the fluid flow, and it shifts the capture interval
asymmetrically on the observer screen.

\begin{figure}[tbhp]
    \centering
    \includegraphics[width=0.43\textwidth]{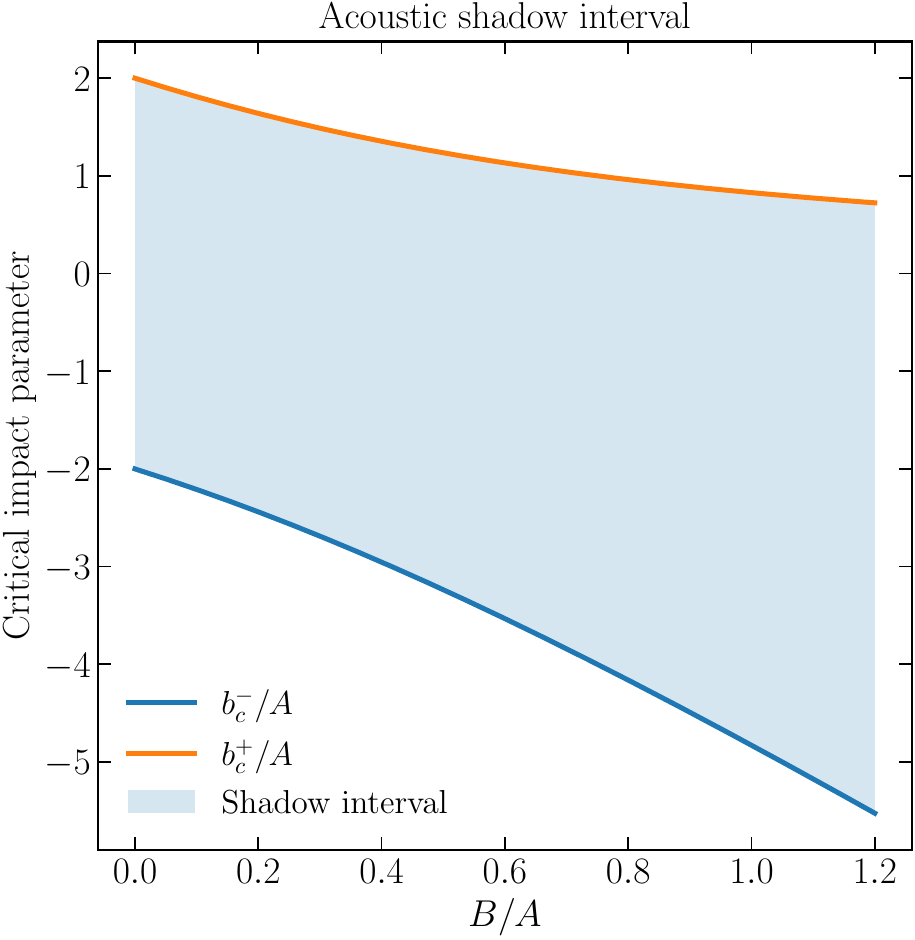}
    \caption{Acoustic shadow interval as a function of $B/A$. The curves are
    $b_c^-/A$ and $b_c^+/A$, and the shaded region denotes impact parameters
    associated with captured rays. The circulation shifts the interval toward
    negative impact parameters and increases the separation between the two
    critical branches.}
    \label{fig:shadow_interval}
\end{figure}

This displacement is quantified in Fig.~\ref{fig:shadow_interval}.  When
$B=0$ the interval is symmetric, $b_c^\pm/A=\pm2$.  For $B\neq0$,
Eqs.~\eqref{eq:shadow_midpoint_width}--\eqref{eq:shadow_midpoint_width_dimensionful} give an exact centroid shift $b_{\rm mid}=-2B$, or $\bar b_{\rm mid}=-2B/A$, and a width $\Delta b=4A\sqrt{1+(B/A)^2}$.  At small circulation the centroid shift is
first order in $B/A$, whereas the width correction is second order.  Thus the
centroid is the cleanest weak-rotation geometric observable.  These statements
are independent of emissivity, transfer index, source velocity and detector
regularization.

\begin{figure*}[tbhp]
    \centering
    \includegraphics[width=0.82\textwidth]{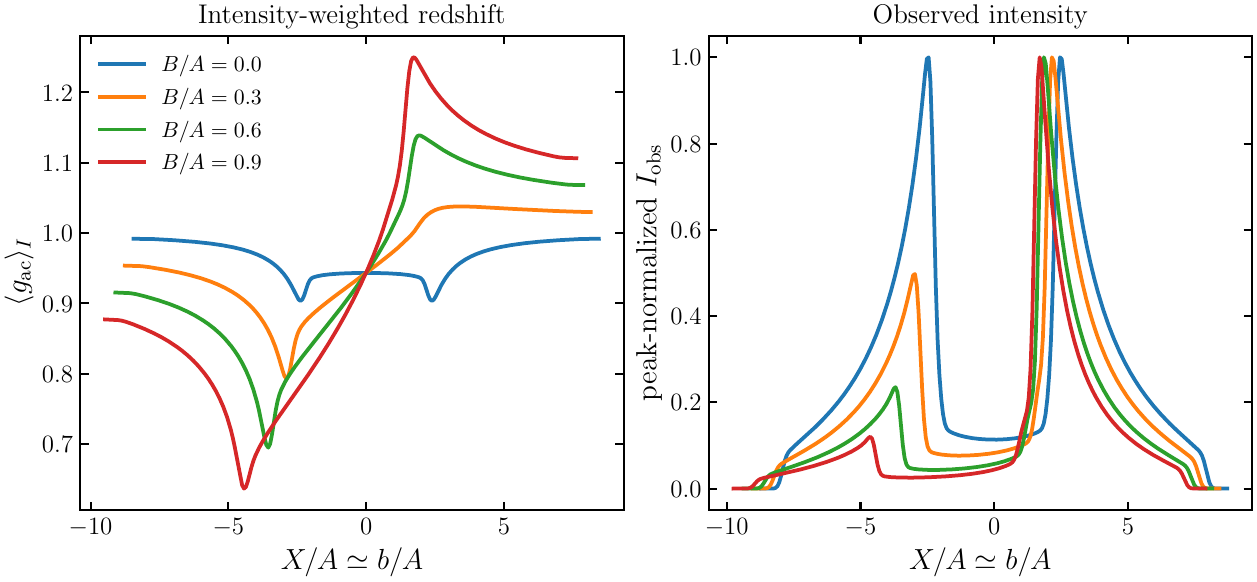}
    \caption{Direct ray-traced extended-disk transfer profiles for different circulation
    parameters.  The upper panel shows the intensity-weighted acoustic
    redshift $\langle g_{\rm ac}\rangle_I$, while the lower panel shows the
    observed intensity normalized by the peak of each curve.  Each intensity
    profile is normalized by its own maximum; the lower panel therefore
    compares morphology and left-right asymmetry, not absolute total flux.}
    \label{fig:fullrt_disk_profiles}
\end{figure*}

Figure~\ref{fig:fullrt_disk_profiles} is the main ray-traced transfer
diagnostic.  The upper panel shows the intensity-weighted redshift profile of
an extended disk, while the lower panel shows the corresponding observed
intensity.  For $B/A=0$ the profiles are symmetric, as required by the
nonrotating geometry. The residual radial structure at $B=0$ is not a rotational asymmetry; it reflects the finite radial extent of the disk and the weighting of different emission radii along the ray. As $B/A$ grows, the redshift distribution develops a
clear left-right tilt: the co-rotating and counter-rotating sides of the image
sample different combinations of acoustic frame dragging, source velocity and
ray bending.  The intensity responds more strongly than the redshift because
$I_{\rm obs}$ contains the factor $g_{\rm ac}^{\eta}$ as well as the
source-to-screen mapping.  The direct ray-tracing transfer calculation is important here: the plotted
curves are not obtained from the approximate ring relation
$\phi_{\rm em}\simeq\arcsin(b/r_s)$, but from real intersections between rays
and the emitting region, followed by direct integration through the disk.

The normalization in the lower panel of Fig.~\ref{fig:fullrt_disk_profiles}
should be read carefully.  Since each curve is divided by its own peak, the
figure emphasizes shape, displacement and branch asymmetry rather than absolute
power.  This choice is useful for comparing the morphology across different
circulations, but it deliberately removes information about the total flux.
That information is instead captured by the branch-integrated quantities
defined in Eqs.~\eqref{eq:branch_fluxes}--\eqref{eq:AI_flux}.  The important
physical message is that increasing $B/A$ does not merely translate the image:
it changes the relative weight of the two sides of the observed profile.

\begin{figure}[tbhp]
    \centering
    \includegraphics[width=0.43\textwidth]{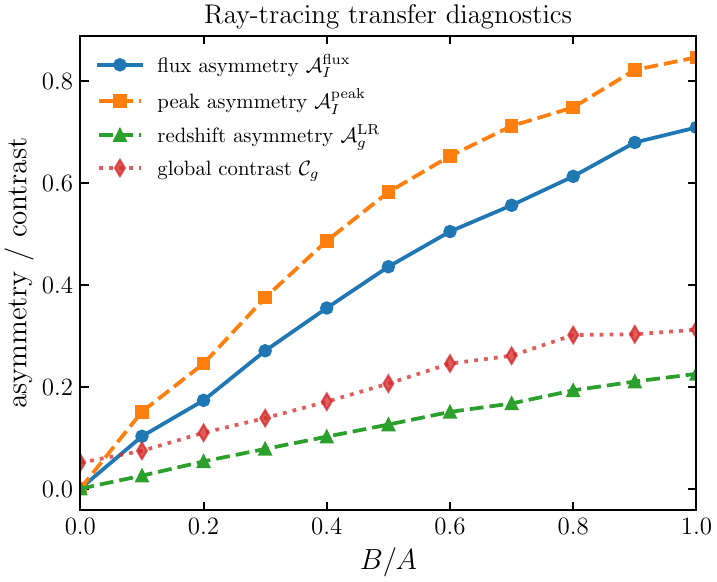}
    \caption{Differential ray-tracing transfer observables as functions of the
    circulation parameter.  The flux asymmetry $\mathcal A_I^{\rm flux}$ is
    the most robust brightness diagnostic, while the peak asymmetry
    $\mathcal A_I^{\rm peak}$ is more sensitive to local magnification.
    The left-right redshift asymmetry $\mathcal A_g^{\rm LR}$ vanishes at
    $B=0$ by symmetry.  The global redshift contrast $\mathcal C_g$ need not
    vanish at $B=0$ because an extended disk contains emission from a range of
    radii with different acoustic redshifts.}
    \label{fig:fullrt_diagnostics}
\end{figure}

Figure~\ref{fig:fullrt_diagnostics} condenses the direct ray-traced profiles into
four scalar observables.  The flux asymmetry $\mathcal A_I^{\rm flux}$ grows
monotonically with $B/A$ and is the preferred brightness diagnostic because it
uses integrated branch fluxes rather than local maxima.  The peak asymmetry
$\mathcal A_I^{\rm peak}$ is larger, as expected: peaks are sensitive to
localized magnification and residual caustic structure.  This does not make the
peak asymmetry unphysical, but it makes it less robust than
$\mathcal A_I^{\rm flux}$.

The redshift diagnostics in Fig.~\ref{fig:fullrt_diagnostics} play two
different roles.  For the axisymmetric fiducial source, the left-right quantity
$\mathcal A_g^{\rm LR}$ is a direct rotational asymmetry and therefore tends
to zero at $B=0$.  It is the direct
redshift analogue of the branch flux asymmetry.  By contrast, the global
contrast $\mathcal C_g$ measures the width of the redshift distribution in the
extended disk.  It remains nonzero even in the nonrotating case because rays
receive contributions from different radii, and different radii have different
acoustic redshifts.  Separating these two redshift measures is essential: a
nonzero global spread in $g_{\rm ac}$ is not, by itself, evidence of rotation,
whereas a nonzero $\mathcal A_g^{\rm LR}$ is.

The diagnostic hierarchy suggested by Fig.~\ref{fig:fullrt_diagnostics} is
therefore clear.  The shadow centroid gives a purely geometric estimate of
$B/A$.  The left-right redshift asymmetry tests whether the frequency shift is
consistent with that circulation.  The flux asymmetry shows how strongly the
same rotation appears in the measured intensity.  The peak asymmetry is useful
as a high-contrast indicator, but should not be used as the only quantitative
observable in the presence of caustics.

\begin{figure}[tbhp]
    \centering
    \includegraphics[width=0.43\textwidth]{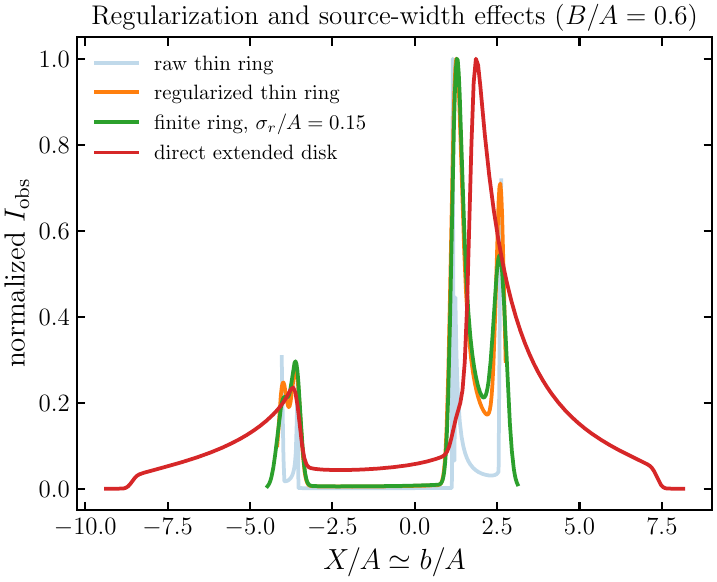}
    \caption{Regularization and source-width effects at $B/A=0.6$.  The raw
    thin-ring limit is dominated by sharp caustic magnification.  A regularized
    thin ring and a finite-width ring suppress the singular behavior, while the
    direct extended-disk calculation gives the physical profile used for the
    main diagnostics. The remaining narrow enhancement in the direct-disk curve is finite after disk integration and detector convolution; it is not the singular thin-ring Jacobian peak.}
    \label{fig:regularization_comparison}
\end{figure}

Figure~\ref{fig:regularization_comparison} explains why the present version of
the paper no longer uses the old thin-ring intensity profiles as final
observables.  A delta-function ring is mathematically useful because it exposes
the source-to-screen map and its Jacobian, but it also produces strong caustic
peaks where $\mathcal J^{-1}$ becomes large.  These peaks are real features of
the idealized infinitesimal-source limit, but they are not robust observables
of a finite experimental source.  The regularized thin-ring curve shows the
effect of capping the caustic magnification, while the finite-width ring shows
how a small physical radial width smooths the profile.  The direct extended
disk is smoother still and is the profile used for the main ray-traced transfer
diagnostics. The remaining narrow enhancement in that curve is finite after disk integration and detector convolution; it is not the singular thin-ring Jacobian peak.

This comparison is methodological rather than merely aesthetic.  It shows that
the old semi-analytic ring figures should not be retained as final evidence for
the brightness asymmetry.  Their qualitative message was correct---rotation
creates a left-right imbalance---but the direct ray-traced disk-transfer calculation is a
better and more defensible observable for the final presentation.

\begin{figure}[tbhp]
    \centering
    \includegraphics[width=0.43\textwidth]{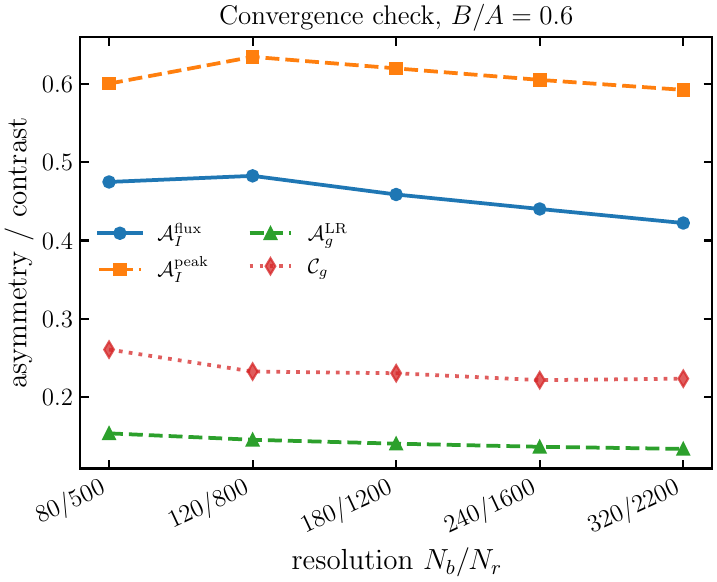}
    \caption{Resolution study for the ray-traced direct-disk observables
    at $B/A=0.6$.  The label $N_b/N_r$ denotes the number of screen points and
    radial integration points.  The two redshift observables and the peak
    asymmetry are stable at the highest resolutions; the flux asymmetry changes
    by a few percent between the last two resolutions.}
    \label{fig:convergence_check}
\end{figure}

Figure~\ref{fig:convergence_check} provides the numerical validation of the
adopted resolution.  Between the two highest resolutions used in the production
runs, the relative changes are approximately $4.3\%$ for
$\mathcal A_I^{\rm flux}$, $2.2\%$ for $\mathcal A_I^{\rm peak}$, $2.0\%$ for
$\mathcal A_g^{\rm LR}$, and below $1\%$ for $\mathcal C_g$.  The flux
asymmetry is the slowest to converge because it integrates the tails of the
profile, where the finite screen sampling and detector convolution matter most.
The redshift asymmetry and global redshift contrast are more stable.  These
numbers are sufficiently controlled for the level of geometric-acoustics
modeling pursued here, and they should be quoted when comparing the predicted
contrast levels with experiment.

\subsection{Robustness under transfer and source choices}
\label{subsec:robustness}

The fiducial profiles in Figs.~\ref{fig:fullrt_disk_profiles}--\ref{fig:fullrt_diagnostics}
should be understood as a controlled baseline rather than as a universal
prediction for every acoustic source.  From the viewpoint of effective-metric
gravity, the important question is not whether a single brightness ratio is
universal, but which observables are geometric, which are transfer-dependent,
and which are most exposed to source degeneracies.  We therefore organize the
robustness discussion around three controlled deformations of the fiducial
model: the transfer exponent, the intrinsic azimuthal emissivity, and the
definition of the left-right split.

\subsubsection{Transfer exponent}
\label{subsubsec:eta_robustness}

The transfer index $\eta$ controls how a fixed redshift field is converted into
a measured intensity contrast.  In the notation of Eq.~\eqref{eq:iobs_general_sum},
the same ray-traced map can be reweighted as
\begin{equation}
 I_{\rm obs}^{(\eta)}(X)=
 \sum_{\rm branches}
 \int g_{\rm ac}^{\eta}
 \mathcal I_{\rm em}[r(\lambda),\phi(\lambda)]W[r(\lambda)]\,d\lambda .
 \label{eq:iobs_eta_family}
\end{equation}
Increasing $\eta$ enhances the bright branch and increases
$\mathcal A_I^{\rm flux}$ and $\mathcal A_I^{\rm peak}$, whereas decreasing
$\eta$ weakens the brightness contrast.  By contrast, the geometric shadow
centroid and the ray-level redshift map are not created by the choice of
$\eta$.  The fiducial choice $\eta=3$ is therefore best viewed as a
high-contrast reference convention, while $\eta=2$ is the natural acoustic
energy-flux reference for an ideal two-dimensional field.  Table~\ref{tab:robustness_eta}
summarizes the expected response of the diagnostics under the standard
$\eta=2,3,4$ comparison.  In a specific experiment, the entries of this table
should be replaced by calibrated detector-dependent values obtained from the
same ray-tracing pipeline.

\begin{table}[tbhp]
\caption{Robustness of the diagnostic hierarchy under changes of the transfer
exponent.  The symbols indicate the qualitative response at fixed geometry and
source model; numerical values are detector-convention dependent.}
\label{tab:robustness_eta}
\begin{ruledtabular}
\begin{tabular}{lcccc}
Case & $\mathcal A_I^{\rm flux}$ & $\mathcal A_I^{\rm peak}$ &
$\mathcal A_g^{\rm LR}$ & $b_{\rm mid}$ \\
\hline
$\eta=2$ & weaker & weaker & unchanged at ray level & unchanged \\
$\eta=3$ & fiducial & fiducial & fiducial weighting & unchanged \\
$\eta=4$ & stronger & stronger & slightly reweighted average & unchanged
\end{tabular}
\end{ruledtabular}
\end{table}

\subsubsection{Intrinsic azimuthal emissivity}
\label{subsubsec:emissivity_robustness}

The intrinsic angular emissivity can imitate part of a brightness asymmetry.
For this reason we distinguish the geometric and frequency-shift diagnostics
from the intensity-only diagnostics.  A useful one-parameter deformation of the
fiducial source is
\begin{equation}
 \mathcal P(\phi)=1+\epsilon_1\cos(\phi-\phi_1),
 \qquad |\epsilon_1|<1 ,
 \label{eq:dipolar_emissivity_test}
\end{equation}
which represents a dipolar hot-spot or density-gradient component.  This
modulation changes $\mathcal A_I^{\rm flux}$ and
$\mathcal A_I^{\rm peak}$ already at fixed geometry, and can therefore
partially mask or mimic the circulation-induced brightness imbalance.  It does
not, however, move the analytic capture interval
$b_c^\pm$ or the centroid $b_{\rm mid}$, and it cannot by itself generate the
same branch-dependent redshift pattern unless it is correlated with the
emitter velocity field.  The robust inference strategy is consequently to fit
or marginalize over $\mathcal P(\phi)$ while requiring consistency among the
shadow centroid, the sign of the redshift tilt, and the flux imbalance.

\subsubsection{Laboratory split versus centroid-corrected split}
\label{subsubsec:split_robustness}

Because the capture interval is displaced when $B\ne0$, the definition of a
left-right asymmetry is not unique.  The laboratory split used in the main
figures compares the two sides of the detector,
\begin{equation}
 F_\pm^{\rm lab}=\int_{\pm X>0}I_{\rm obs}(X)dX,
 \label{eq:lab_split_flux}
\end{equation}
whereas a centroid-corrected split removes the pure geometric displacement
before comparing the two branches,
\begin{equation}
 F_\pm^{\rm cen}=\int_{\pm (X-b_{\rm mid})>0}I_{\rm obs}(X)dX .
 \label{eq:centroid_split_flux}
\end{equation}
The first convention is closest to what a fixed detector records; the second
is closer to a transfer diagnostic after the geometric shift has been fitted.
Both are meaningful, but they answer different questions.  A strong
circulation interpretation should not rely on only one convention.  It should
show that the inferred value of $B/A$ from the centroid is compatible with the
residual asymmetries extracted after the centroid displacement is removed.
Table~\ref{tab:source_split_robustness} summarizes the qualitative diagnostic
response to intrinsic emissivity modulation and to the two left-right split
conventions.  The entries are intended as a robustness guide for the same
ray-tracing pipeline; numerical values depend on the adopted source model,
detector convolution, and transfer exponent.

\begin{table*}[tbhp]
\caption{Qualitative response of the main diagnostics to source and split
choices at fixed acoustic geometry.  Entries classify robustness and do not
replace experiment-specific calibration.}
\label{tab:source_split_robustness}
\begin{ruledtabular}
\footnotesize
\setlength{\tabcolsep}{5pt}
\begin{tabular}{lllll}
Case & $b_{\rm mid}$ & $\mathcal A_g^{\rm LR}$ &
$\mathcal A_I^{\rm flux}$ & $\mathcal A_I^{\rm peak}$ \\
\hline
Axisymmetric source & reference & direct tilt & fiducial imbalance & caustic sensitive \\
Dipolar $\mathcal P(\phi)$ & unchanged & reweighted & source degenerate & source sensitive \\
Lab split $X=0$ & measured & left--right average & detector-side flux & detector-side peak \\
Centroid split & subtracted & residual contrast & corrected flux & residual peak
\end{tabular}
\end{ruledtabular}
\end{table*}

\subsection{Effective-metric and laboratory interpretation}
\label{subsec:lab_interpretation}

The hierarchy above clarifies how the calculation should be used in an
effective-metric or laboratory setting.  The geometric-acoustics description
requires wavelengths short compared with the horizon scale and with the
distance between the dominant emitting region and the horizon.  Dissipation,
dispersion, turbulence and finite depth mainly smooth the sharpest
near-critical features and reduce the contrast of high-order winding rays.
They are therefore expected to affect peak observables before they invalidate
the centroid or branch-integrated diagnostics.  Table~\ref{tab:lab_requirements}
summarizes the main requirements and the corresponding observable most
affected.

\begin{table*}[tbhp]
\caption{Practical requirements for applying the ray-tracing diagnostics to a
rotating acoustic effective metric or laboratory vortex-flow analogue.  The
entries are scaling criteria, not universal numerical thresholds.}
\label{tab:lab_requirements}
\begin{ruledtabular}
\begin{tabular}{lll}
Effect or requirement & Controlled by & Main observable affected \\
\hline
Geometric-acoustics validity & $\lambda/A\ll1$ and $\lambda/(r_{\rm em}-r_h)\ll1$ & Shadow edge and near-critical peaks \\
Dispersion & Medium dispersion relation & Frequency-dependent shadow smearing \\
Viscous attenuation & Damping length compared with ray path length & High-order winding rays and peak contrast \\
Finite source width & $\sigma_r/A$ or disk thickness & Caustic regularization and peak asymmetry \\
Detector/reconstruction resolution & $\sigma_X/A$ and screen sampling & Flux tails and branch-integrated asymmetry \\
Intrinsic emissivity asymmetry & $\mathcal P(\phi)$ or hot spots & Brightness asymmetry and peak location \\
Finite-depth corrections & Depth-to-radius expansion parameters & Effective metric and calibrated values of $A,B$ \\
Turbulent velocity fluctuations & Background-flow noise level & Redshift-map scatter and signal-to-noise
\end{tabular}
\end{ruledtabular}
\end{table*}

\paragraph{Relation to semi-analytic transfer prescriptions.}
Semi-analytic transfer prescriptions based on approximate source angles or on $\mathcal J=1$ are useful for developing intuition, but they are not used as final observables here. Such prescriptions correctly indicate that emitter kinematics, emissivity and transfer exponent matter; however, the figures that support the final claims use direct ray tracing, real ray-source intersections, finite source width or direct disk integration, and explicit convergence checks. This is why the final analysis is based on the ray-traced transfer profiles and diagnostics in Figs.~\ref{fig:fullrt_disk_profiles}--\ref{fig:convergence_check}.

\paragraph{Geometric-acoustics approximation and robustness.}
The framework operates in the geometric-acoustics limit, which is
self-consistent when the acoustic wavelength satisfies $\lambda\ll A$ and
$\lambda\ll r_s-r_h$ for the dominant emitting region.  Real acoustic media are
dispersive at short wavelengths.  Dispersive corrections modify near-critical
rays and smear the shadow edge over a frequency-dependent width.  In simple geometric-optics estimates near a turning point, corrections may scale with fractional powers of $\lambda/A$, for example as $(\lambda/A)^{2/3}$, although the exponent and coefficient are model dependent \cite{Barcelo2011}.  Viscous dissipation attenuates rays with long
path length and therefore suppresses high-order near-critical orbits.  Turbulent
velocity fluctuations add stochastic phase and amplitude noise, limiting the
signal-to-noise ratio of the asymmetry measurements.  Finite tank depth adds
metric corrections that are expected to scale with powers of the depth-to-radius
ratio, for example $(h/r)^2$ in shallow-depth expansions.  These effects do not
remove the leading-order picture---a displaced shadow, a left-right redshift
asymmetry, and a flux imbalance correlated with the circulation---but they set
the practical requirements for laboratory implementation.

Taken together, Figs.~\ref{fig:null_rays}--\ref{fig:convergence_check} summarize the geometric, transfer, regularization, and convergence aspects of the calculation.  The first two figures establish the
source-independent geometry.  The direct ray-traced disk profiles show the
observable redshift and intensity distributions.  The differential diagnostics
show which quantities best measure rotation.  The regularization comparison
explains why the raw thin-ring limit is not a final physical observable.  The
convergence study quantifies the numerical stability of the adopted procedure.

\section{Conclusions}
\label{sec:conclusions}

We have developed an impact-parameter-resolved ray-tracing transfer framework
for the acoustic shadow, redshift distribution, and observed flux profile of a
rotating effective black-hole geometry.  The model is the draining-bathtub
spacetime: the draining parameter $A$ sets the acoustic horizon, while the
circulation parameter $B$ generates an ergoregion and hydrodynamic frame
dragging.  The null-ray structure yields analytic critical impact parameters,
Eq.~\eqref{eq:critical_impact_explicit}, and therefore exact expressions for
the acoustic shadow centroid and width,
\begin{equation}
 b_{\rm mid}=-2B,\qquad
 \Delta b=4A\sqrt{1+(B/A)^2}.
\end{equation}
These quantities are geometric and do not depend on the source emissivity,
transfer exponent, or detector normalization.

The main advance over a semi-analytic ring prescription is that the observable
profiles are computed with real ray-source intersections and with an explicit
source-to-screen mapping.  The raw thin-ring limit remains useful because it
exposes the caustic structure of the map, but it is not a robust
finite-resolution observable.  Physical profiles require regularization,
finite source width, detector convolution, or direct integration through an
extended emitting disk.  We use the latter as the main observable and retain
the ring calculation only as a diagnostic limiting case.

The resulting hierarchy of observables is directly analogous to the logic of
shadow-based rotation diagnostics in gravitational ray tracing.  The shadow
centroid provides the cleanest geometric estimate of circulation.  The
left-right redshift asymmetry $\mathcal A_g^{\rm LR}$ tests whether the
frequency shift transferred along the rays is compatible with that circulation.
The branch-integrated flux asymmetry $\mathcal A_I^{\rm flux}$ measures how
strongly the same rotation appears in the observed intensity and is less
sensitive to caustic peaks than $\mathcal A_I^{\rm peak}$.  The global redshift
contrast $\mathcal C_g$ has a different interpretation: it measures the spread
of redshifts across the emitting disk and can be nonzero even in a nonrotating
configuration.  Separating these observables is essential, because neither a
brightness imbalance nor a redshift spread is by itself a unique signature of
rotation.

We have also identified the main degeneracies that must be controlled before a
circulation estimate can be regarded as robust.  The transfer exponent $\eta$
depends on the measured acoustic quantity and detector response; intrinsic
azimuthal emissivity can imitate part of a flux asymmetry; and the laboratory
left-right split differs from a centroid-corrected split after the geometric
shadow displacement has been removed.  The safest inference strategy is
therefore multi-observable: fit the geometric centroid, check the sign and
magnitude of the redshift tilt, and then interpret the flux imbalance after
marginalizing over the source emissivity and transfer convention.

The convergence study gives a numerical baseline for the reported contrast
levels.  At $B/A=0.6$, the two highest resolutions change
$\mathcal A_I^{\rm flux}$ by about $4\%$,
$\mathcal A_I^{\rm peak}$ and $\mathcal A_g^{\rm LR}$ by about $2\%$, and
$\mathcal C_g$ by less than $1\%$ within the adopted smoothing and screen
sampling conventions.  Wave simulations will be needed to quantify dispersive,
dissipative, and finite-depth corrections beyond the ray limit, but these
effects are expected first to smooth the sharpest near-critical features rather
than to erase the leading geometric displacement of the acoustic shadow.

The framework therefore provides a controlled analogue-gravity setting for
transfer observables in rotating black-hole effective metrics.  Future
extensions should include frequency-dependent dispersive rays, viscous damping
along long path-length trajectories, finite-depth corrections to the effective
metric, time-dependent or turbulent vortex backgrounds, explicit marginalization
over nonaxisymmetric source emissivity, and full two-dimensional image-plane
reconstruction beyond the impact-parameter-resolved profiles considered here.

\begin{acknowledgments}
F. Ahmed gratefully acknowledges the Inter University Center for Astronomy and Astrophysics (IUCAA), Pune, India, for the conferment of a visiting associateship. F. M. Belchior would like to express gratitude to the Conselho Nacional de Desenvolvimento Cient\'{i}fico e Tecnol\'{o}gico CNPq for grant No. 151845/2025-5. E. O. Silva acknowledges the support from Conselho Nacional de Desenvolvimento Cient\'{i}fico e Tecnol\'{o}gico (CNPq) (grants 306308/2022-3), Funda\c c\~ao de Amparo \`{a} Pesquisa e ao Desenvolvimento Cient\'{i}fico e Tecnol\'{o}gico do Maranh\~ao (FAPEMA) (grants UNIVERSAL-06395/22), and Coordena\c c\~ao de Aperfei\c coamento de Pessoal de N\'{i}vel Superior (CAPES) - Brazil (Code 001).
\end{acknowledgments}

\section*{Data Availability Statement}
No experimental data were used in this work.  The numerical data and routines
used to generate the figures and diagnostic curves are available from the
corresponding author upon reasonable request.

\appendix

\section{Hamiltonian formulation of the acoustic rays}
\label{app:inverse_metric}

The main text uses the first-integral form of the null-ray equations because it
is compact and makes the critical impact parameters transparent.  For numerical
ray tracing, however, it is often useful to recast the same problem in
Hamiltonian form.  This appendix gives that formulation and clarifies how it is
related to the screen-to-source ray tracing used in the figures.

The inverse metric associated with Eq.~\eqref{eq:bl_metric} has nonzero
components
\begin{align}
& g^{tt}=\frac{1}{f(r)},
 \qquad
 g^{t\phi}=\frac{B}{r^2 f(r)},
 \label{eq:inverse_metric_t}\\
& g^{rr}=-f(r),
 \qquad
 g^{\phi\phi}=-\frac{g(r)}{r^2 f(r)}.
 \label{eq:inverse_metric_spatial}
\end{align}
The acoustic rays are the null characteristics of the effective metric and may
therefore be generated from the Hamiltonian
\begin{equation}
 \mathcal H(x^\mu,k_\mu)=\frac12 g^{\mu\nu}k_\mu k_\nu=0.
 \label{eq:hamiltonian}
\end{equation}
Explicitly,
\begin{equation}
2\mathcal H=
\frac{k_t^2}{f(r)}
+\frac{2B}{r^2 f(r)}k_t k_\phi
-f(r)k_r^2
-\frac{g(r)}{r^2 f(r)}k_\phi^2 .
\label{eq:hamiltonian_explicit}
\end{equation}
The Hamilton equations are
\begin{equation}
 \frac{dx^\mu}{d\lambda}=\frac{\partial\mathcal H}{\partial k_\mu},
 \qquad
 \frac{dk_\mu}{d\lambda}=-\frac{\partial\mathcal H}{\partial x^\mu}.
 \label{eq:hamilton_eqs}
\end{equation}
Because the metric is independent of $t$ and $\phi$, the quantities
\begin{equation}
 k_t=E,
 \qquad
 k_\phi=-L
 \label{eq:hamilton_conserved}
\end{equation}
are conserved along the ray.  This convention is the same as that used in the
main text: the impact parameter is $b=L/E$, while the covariant angular
momentum is $k_\phi=-L$.

Using Eq.~\eqref{eq:hamilton_eqs}, one obtains
\begin{equation}
 \dot t=\frac{\partial\mathcal H}{\partial k_t}
 =\frac{E}{f(r)}-\frac{BL}{r^2 f(r)}
 =\frac{E-BL/r^2}{f(r)},
 \label{eq:app_tdot}
\end{equation}
and
\begin{equation}
 \dot\phi=\frac{\partial\mathcal H}{\partial k_\phi}
 =\frac{BE}{r^2 f(r)}+\frac{g(r)L}{r^2 f(r)}.
 \label{eq:app_phidot_intermediate}
\end{equation}
Using $g(r)=f(r)-B^2/r^2$, this can be rewritten as
\begin{equation}
 \dot\phi=
 \frac{L}{r^2}+\frac{BE-LB^2/r^2}{r^2 f(r)},
 \label{eq:app_phidot}
\end{equation}
which is Eq.~\eqref{eq:phidot} of the main text.  Thus the Hamiltonian and
first-integral approaches are exactly equivalent.

The radial equation follows from the null constraint
$\mathcal H=0$.  With $k_t=E$ and $k_\phi=-L$, one finds
\begin{equation}
 f(r)^2 k_r^2
 =
 E^2-\frac{b^2g(r)+2Bb}{r^2}E^2,
 \label{eq:app_kr}
\end{equation}
or, equivalently,
\begin{equation}
 \frac{\dot r^2}{E^2}
 =
 1-\frac{b^2g(r)+2Bb}{r^2}.
 \label{eq:app_radial}
\end{equation}
This is Eq.~\eqref{eq:radial_b}.  The sign of $\dot r$ distinguishes the inward
and outward parts of a scattered ray.  In the numerical implementation this
distinction is essential, because the same screen impact parameter can
intersect the emitting region once on the inward branch and again after a
turning point on the outward branch.

The Hamiltonian viewpoint is useful for three reasons.  First, it makes clear
that the calculation is a characteristic calculation for the acoustic wave
equation in the geometric-acoustics limit.  Second, it provides a direct route
to more general ray tracing, for example when one includes dispersive
corrections, time-dependent background flows, or noncircular source regions.
Third, it gives a clean way of defining the covariant radial momentum used in
the general redshift formula.  Since
\begin{equation}
 k_r=g_{rr}\dot r=-\frac{\dot r}{f(r)},
 \label{eq:app_kr_covariant}
\end{equation}
we have
\begin{equation}
 \frac{k_r}{E}
 =
 -\frac{1}{f(r)}
 \frac{\dot r}{E}
 =
 \mp\frac{\sqrt{F(r;b,B)}}{f(r)},
 \label{eq:app_kr_over_E}
\end{equation}
where
\begin{equation}
F(r;b,B)=1-\frac{b^2g(r)+2Bb}{r^2}.
\label{eq:app_F}
\end{equation}
The upper/lower sign depends on whether the ray is on the outward or inward
branch.  This is the quantity that enters Eq.~\eqref{eq:gac_general} when an
emitter has a nonzero radial velocity $v^r_{\rm em}$.

In the present paper we used the reduced first-integral form for the production
figures, because the spacetime is stationary and axisymmetric.  Nevertheless,
the Hamiltonian formulation above is the natural starting point for future
extensions in which the vortex flow is time dependent, the effective metric
contains additional corrections, or the full two-dimensional image plane is
ray-traced without reducing the problem to the effective screen coordinate
$X\simeq b$.

\section{Dimensionless variables and numerical scaling}
\label{app:dimensionless}

The numerical work is performed in units in which the draining parameter is set
to unity.  This is not a loss of generality.  The parameter $A$ fixes the
horizon radius and therefore provides the natural length scale of the
problem.  We define
\begin{equation}
 \bar r=\frac{r}{A},
 \qquad
 \bar B=\frac{B}{A},
 \qquad
 \bar b=\frac{b}{A},
 \qquad
 \bar X=\frac{X}{A}.
 \label{eq:dimensionless_variables}
\end{equation}
In these variables the horizon and ergosurface are located at
\begin{equation}
 \bar r_h=1,
 \qquad
 \bar r_e=\sqrt{1+\bar B^2},
 \label{eq:dimensionless_ergo}
\end{equation}
and the metric functions become
\begin{equation}
 f(\bar r)=1-\frac{1}{\bar r^2},
 \qquad
 g(\bar r)=1-\frac{1+\bar B^2}{\bar r^2}.
 \label{eq:dimensionless_fg}
\end{equation}
All figures in the main text use this normalization.

The usefulness of this scaling is that the circulation appears only through the
dimensionless ratio $\bar B=B/A$.  The critical impact parameters are therefore
\begin{equation}
 \bar b_c^\pm=-2\bar B\pm2\sqrt{1+\bar B^2},
 \label{eq:dimensionless_bc}
\end{equation}
while the centroid and width of the shadow interval are
\begin{equation}
 \bar b_{\rm mid}=-2\bar B,
 \qquad
 \Delta \bar b=4\sqrt{1+\bar B^2}.
 \label{eq:dimensionless_centroid_width}
\end{equation}
Thus, once the results have been computed for $A=1$, they can be restored to
physical units by multiplying all lengths and impact parameters by $A$.  For
example,
\begin{equation}
 b_c^\pm=A\bar b_c^\pm,
 \qquad
 X=A\bar X,
 \qquad
 r=A\bar r .
 \label{eq:restore_units}
\end{equation}

The same scaling applies to the source model.  The fiducial extended disk used
in the direct ray-tracing figures has
\begin{equation}
 \bar r_{\rm in}=2,
 \qquad
 \bar r_{\rm out}=8,
 \qquad
 \mathcal R(\bar r)\propto \bar r^{-p},
 \qquad
 p=2.
 \label{eq:dimensionless_disk}
\end{equation}
The finite-width ring used only for the methodological comparison in Fig.~\ref{fig:regularization_comparison} has
\begin{equation}
 \bar\sigma_r=\frac{\sigma_r}{A}=0.15,
 \label{eq:dimensionless_sigma_r}
\end{equation}
and the detector smoothing scale is
\begin{equation}
 \bar\sigma_X=\frac{\sigma_X}{A}=0.12.
 \label{eq:dimensionless_sigma_X}
\end{equation}
These two quantities are not meant to represent universal physics.  They encode
the finite radial width of the source and the finite resolution of the
detector/reconstruction procedure.  Their role is to regularize the ideal
geometric-acoustics caustics and to turn the ray map into an observable
finite-resolution intensity profile.

The numerical resolution is specified by two integers.  The screen resolution
$N_b$ determines how finely the impact parameter is sampled, and the radial
resolution $N_r$ determines how finely each ray is integrated.  In the fiducial
runs we use
\begin{equation}
 N_b=260,
 \qquad
 N_r=1300,
 \label{eq:dimensionless_resolution}
\end{equation}
and test convergence by increasing both quantities. The fiducial values are used for the production profiles, while the convergence test increases them up to $N_b=320$ and $N_r=2200$. The convergence plot in
Fig.~\ref{fig:convergence_check} should therefore be interpreted as a test of
the complete observable pipeline: ray integration, disk integration, detector
convolution, and extraction of the asymmetry diagnostics.

Finally, all intensities in the figures are normalized by an arbitrary source
amplitude $I_0$.  The plotted intensity profiles are therefore dimensionless.
When a profile is divided by its own maximum, as in
Fig.~\ref{fig:fullrt_disk_profiles}, the purpose is to compare morphology,
left-right asymmetry and peak location, not absolute power.  Absolute
branch-integrated information is instead retained in the fluxes $F_\pm$ and in
the asymmetry $\mathcal A_I^{\rm flux}$.

%


\end{document}